
%
\def\unlockat{\catcode`\@=11}
\def\lockat{\catcode`\@=12}
\unlockat
\def\d@f@ult{} \newif\ifamsfonts \newif\ifafour
\def\m@ssage{\immediate\write16}  \m@ssage{}
\m@ssage{hep-th preprint macros.  Last modified 16/10/92 (jmf).}
\message{These macros work with AMS Fonts 2.1 (available via ftp from}
\message{e-math.ams.com).  If you have them simply hit "return"; if}
\message{you don't, type "n" now: }
\endlinechar=-1  
\read-1 to\@nswer
\endlinechar=13
\ifx\@nswer\d@f@ult\amsfontstrue
    \m@ssage{(Will load AMS fonts.)}
\else\amsfontsfalse\m@ssage{(Won't load AMS fonts.)}\fi
\message{The default papersize is A4.  If you use US 8.5" x 11"}
\message{type an "a" now, else just hit "return": }
\endlinechar=-1  
\read-1 to\@nswer
\endlinechar=13
\ifx\@nswer\d@f@ult\afourtrue
    \m@ssage{(Using A4 paper.)}
\else\afourfalse\m@ssage{(Using US 8.5" x 11".)}\fi
\nonstopmode
%
%

\font\twelverm=cmr12
\font\ninerm=cmr9
\font\sixrm=cmr6
\font\fourteenbf=cmbx12 scaled\magstep1
\font\twelvebf=cmbx12
\font\ninebf=cmbx9
\font\sixbf=cmbx6
\font\fourteeni=cmmi12 scaled\magstep1      \skewchar\fourteeni='177
\font\twelvei=cmmi12                        \skewchar\twelvei='177
\font\ninei=cmmi9                           \skewchar\ninei='177
\font\sixi=cmmi6                            \skewchar\sixi='177
\font\fourteensy=cmsy10 scaled\magstep2     \skewchar\fourteensy='60
\font\twelvesy=cmsy10 scaled\magstep1       \skewchar\twelvesy='60
\font\ninesy=cmsy9                          \skewchar\ninesy='60
\font\sixsy=cmsy6                           \skewchar\sixsy='60
\font\fourteenex=cmex10 scaled\magstep2
\font\twelveex=cmex10 scaled\magstep1

\ifamsfonts
   \font\ninex=cmex9
   
   \font\sixex=cmex7 at 6pt
   
\else
   \font\ninex=cmex10 at 9pt
   
   \font\sixex=cmex10 at 6pt
   
\fi
\font\fourteensl=cmsl10 scaled\magstep2
\font\twelvesl=cmsl10 scaled\magstep1

\font\sevensl=cmsl10 at 7pt
\font\sixsl=cmsl10 at 6pt

\font\fourteenit=cmti12 scaled\magstep1
\font\twelveit=cmti12

\font\fourteentt=cmtt12 scaled\magstep1
\font\twelvett=cmtt12
\font\fourteencp=cmcsc10 scaled\magstep2
\font\twelvecp=cmcsc10 scaled\magstep1

\ifamsfonts
   
\else
   
\fi
\newfam\cpfam
\font\fourteenss=cmss12 scaled\magstep1
\font\twelvess=cmss12
\font\tenss=cmss10
\font\niness=cmss9

\font\sevenss=cmss8 at 7pt
\font\sixss=cmss8 at 6pt
\newfam\ssfam
\newfam\msafam \newfam\msbfam \newfam\eufam
\ifamsfonts
 \font\fourteenmsa=msam10 scaled\magstep2
 \font\twelvemsa=msam10 scaled\magstep1
 \font\tenmsa=msam10
 \font\ninemsa=msam9
 \font\sevenmsa=msam7
 \font\sixmsa=msam6
 \font\fourteenmsb=msbm10 scaled\magstep2
 \font\twelvemsb=msbm10 scaled\magstep1
 \font\tenmsb=msbm10
 \font\ninemsb=msbm9
 \font\sevenmsb=msbm7
 \font\sixmsb=msbm6
 \font\fourteeneu=eufm10 scaled\magstep2
 \font\twelveeu=eufm10 scaled\magstep1
 \font\teneu=eufm10
 \font\nineeu=eufm9
 
 \font\seveneu=eufm7
 \font\sixeu=eufm6
 \def\hexnumber@#1{\ifnum#1<10 \number#1\else
  \ifnum#1=10 A\else\ifnum#1=11 B\else\ifnum#1=12 C\else
  \ifnum#1=13 D\else\ifnum#1=14 E\else\ifnum#1=15 F\fi\fi\fi\fi\fi\fi\fi}
 \def\hexmsa{\hexnumber@\msafam}
 \def\hexmsb{\hexnumber@\msbfam} 
\fi
\newdimen\b@gheight             \b@gheight=12pt
\newcount\f@ntkey               \f@ntkey=0
\def\f@m{\afterassignment\samef@nt\f@ntkey=}
\def\samef@nt{\fam=\f@ntkey \the\textfont\f@ntkey\relax}
\def\rm{\f@m0 }
\def\mit{\f@m1 }
\def\cal{\f@m2 }
\def\it{\f@m\itfam}
\def\sl{\f@m\slfam}
\def\bf{\f@m\bffam}
\def\tt{\f@m\ttfam}
\def\caps{\f@m\cpfam}
\def\ssf{\f@m\ssfam}
\ifamsfonts
 \def\msa{\f@m\msafam}
 \def\msb{\f@m\msbfam} \let\bb=\msb
 \def\eu{\f@m\eufam}
\else
 \let \bb=\bf \let\eu=\bf
\fi
\def\fourteenpoint{\relax
    \textfont0=\fourteencp          \scriptfont0=\tenrm
      \scriptscriptfont0=\sevenrm
    \textfont1=\fourteeni           \scriptfont1=\teni
      \scriptscriptfont1=\seveni
    \textfont2=\fourteensy          \scriptfont2=\tensy
      \scriptscriptfont2=\sevensy
    \textfont3=\fourteenex          \scriptfont3=\twelveex
      \scriptscriptfont3=\tenex
    \textfont\itfam=\fourteenit     \scriptfont\itfam=\tenit
    \textfont\slfam=\fourteensl     \scriptfont\slfam=\tensl
      \scriptscriptfont\slfam=\sevensl
    \textfont\bffam=\fourteenbf     \scriptfont\bffam=\tenbf
      \scriptscriptfont\bffam=\sevenbf
    \textfont\ttfam=\fourteentt
    \textfont\cpfam=\fourteencp
    \textfont\ssfam=\fourteenss     \scriptfont\ssfam=\tenss
      \scriptscriptfont\ssfam=\sevenss
    \ifamsfonts
       \textfont\msafam=\fourteenmsa     \scriptfont\msafam=\tenmsa
         \scriptscriptfont\msafam=\sevenmsa
       \textfont\msbfam=\fourteenmsb     \scriptfont\msbfam=\tenmsb
         \scriptscriptfont\msbfam=\sevenmsb
       \textfont\eufam=\fourteeneu     \scriptfont\eufam=\teneu
         \scriptscriptfont\eufam=\seveneu \fi
    \samef@nt
    \b@gheight=14pt
    \setbox\strutbox=\hbox{\vrule height 0.85\b@gheight
                                depth 0.35\b@gheight width\z@ }}
\def\twelvepoint{\relax
    \textfont0=\twelverm          \scriptfont0=\ninerm
      \scriptscriptfont0=\sixrm
    \textfont1=\twelvei           \scriptfont1=\ninei
      \scriptscriptfont1=\sixi
    \textfont2=\twelvesy           \scriptfont2=\ninesy
      \scriptscriptfont2=\sixsy
    \textfont3=\twelveex          \scriptfont3=\ninex
      \scriptscriptfont3=\sixex
    \textfont\itfam=\twelveit    
    \textfont\slfam=\twelvesl    
      \scriptscriptfont\slfam=\sixsl
    \textfont\bffam=\twelvebf     \scriptfont\bffam=\ninebf
      \scriptscriptfont\bffam=\sixbf
    \textfont\ttfam=\twelvett
    \textfont\cpfam=\twelvecp
    \textfont\ssfam=\twelvess     \scriptfont\ssfam=\niness
      \scriptscriptfont\ssfam=\sixss
    \ifamsfonts
       \textfont\msafam=\twelvemsa     \scriptfont\msafam=\ninemsa
         \scriptscriptfont\msafam=\sixmsa
       \textfont\msbfam=\twelvemsb     \scriptfont\msbfam=\ninemsb
         \scriptscriptfont\msbfam=\sixmsb
       \textfont\eufam=\twelveeu     \scriptfont\eufam=\nineeu
         \scriptscriptfont\eufam=\sixeu \fi
    \samef@nt
    \b@gheight=12pt
    \setbox\strutbox=\hbox{\vrule height 0.85\b@gheight
                                depth 0.35\b@gheight width\z@ }}
\twelvepoint
%
%
\baselineskip = 15pt plus 0.2pt minus 0.1pt 
\lineskip = 1.5pt plus 0.1pt minus 0.1pt
\lineskiplimit = 1.5pt
\parskip = 6pt plus 2pt minus 1pt
\interlinepenalty=50
\interfootnotelinepenalty=5000
\predisplaypenalty=9000
\postdisplaypenalty=500
\hfuzz=1pt
\vfuzz=0.2pt
\dimen\footins=24 truecm 
\ifafour
 \hsize=16cm \vsize=22cm
\else
 \hsize=6.5in \vsize=9in
\fi
%
%
\skip\footins=\medskipamount
\newcount\fnotenumber
\def\clearfnotenumber{\fnotenumber=0} \clearfnotenumber
\def\fnote{\global\advance\fnotenumber by1 \generatefootsymbol
 \footnote{$^{\footsymbol}$}}
\def\fd@f#1 {\xdef\footsymbol{\mathchar"#1 }}
\def\generatefootsymbol{\iffrontpage\ifcase\fnotenumber
\or \fd@f 279 \or \fd@f 27A \or \fd@f 278 \or \fd@f 27B
\else  \fd@f 13F \fi
\else\xdef\footsymbol{\the\fnotenumber}\fi}
%
%
\newcount\secnumber \newcount\appnumber
\def\clearappnumber{\appnumber=64} \def\clearsecnumber{\secnumber=0}
\clearsecnumber \clearappnumber
\newif\ifs@c 
\newif\ifs@cd 
\s@cdtrue 
\def\unsectioned{\s@cdfalse\let\section=\subsection}
\newskip\sectionskip         \sectionskip=\medskipamount
\newskip\headskip            \headskip=8pt plus 3pt minus 3pt
\newdimen\sectionminspace    \sectionminspace=10pc
\def\Titlestyle#1{\par\begingroup \interlinepenalty=9999
     \leftskip=0.02\hsize plus 0.23\hsize minus 0.02\hsize
     \rightskip=\leftskip \parfillskip=0pt
     \advance\baselineskip by 0.5\baselineskip
     \hyphenpenalty=9000 \exhyphenpenalty=9000
     \tolerance=9999 \pretolerance=9000
     \spaceskip=0.333em \xspaceskip=0.5em
     \fourteenpoint
  \noindent #1\par\endgroup }
\def\titlestyle#1{\par\begingroup \interlinepenalty=9999
     \leftskip=0.02\hsize plus 0.23\hsize minus 0.02\hsize
     \rightskip=\leftskip \parfillskip=0pt
     \hyphenpenalty=9000 \exhyphenpenalty=9000
     \tolerance=9999 \pretolerance=9000
     \spaceskip=0.333em \xspaceskip=0.5em
     \fourteenpoint
   \noindent #1\par\endgroup }
\def\spacecheck#1{\dimen@=\pagegoal\advance\dimen@ by -\pagetotal
   \ifdim\dimen@<#1 \ifdim\dimen@>0pt \vfil\break \fi\fi}
\def\section#1{\cleareqnumber \s@ctrue \global\advance\secnumber by1
   \par \ifnum\the\lastpenalty=30000\else
   \penalty-200\vskip\sectionskip \spacecheck\sectionminspace\fi
   \noindent {\caps\enspace\S\the\secnumber\quad #1}\par
   \nobreak\vskip\headskip \penalty 30000 }
\def\undertext#1{\vtop{\hbox{#1}\kern 1pt \hrule}}
\def\subsection#1{\par
   \ifnum\the\lastpenalty=30000\else \penalty-100\smallskip
   \spacecheck\sectionminspace\fi
   \noindent\undertext{#1}\enspace \vadjust{\penalty5000}}

\def\appendix#1{\cleareqnumber \s@cfalse \global\advance\appnumber by1
   \par \ifnum\the\lastpenalty=30000\else
   \penalty-200\vskip\sectionskip \spacecheck\sectionminspace\fi
   \noindent {\caps\enspace Appendix \char\the\appnumber\quad #1}\par
   \nobreak\vskip\headskip \penalty 30000 }
\def\ack{\par\penalty-100\medskip \spacecheck\sectionminspace
   \line{\fourteencp\hfil ACKNOWLEDGEMENTS\hfil}%
\nobreak\vskip\headskip }
\def\refs{\begingroup \par\penalty-100\medskip \spacecheck\sectionminspace
   \line{\fourteencp\hfil REFERENCES\hfil}%
\nobreak\vskip\headskip \frenchspacing }
\def\endrefs{\par\endgroup}
%
%
\newif\iffrontpage \frontpagefalse
\headline={\hfil}
\footline={\iffrontpage\hfil\else \hss\twelverm
-- \folio\ --\hss \fi }
%
%
\newskip\frontpageskip \frontpageskip=12pt plus .5fil minus 2pt
\def\titlepage{\global\frontpagetrue\hrule height\z@ \relax
               \pubblock\relax }
\def\endtitlepage{\vfil\break\clearfnotenumber\frontpagefalse}
\def\title#1{\vskip\frontpageskip\Titlestyle{\caps #1}\vskip3\headskip}
\def\author#1{\vskip.5\frontpageskip\titlestyle{\caps #1}\nobreak}
\def\and{\par\kern 5pt \centerline{\sl and}}
\def\andauthor{\vskip.5\frontpageskip\centerline{and}\author}

\def\address#1{\par\kern 5pt\titlestyle{\it #1}}
\def\andaddress{\par\kern 5pt \centerline{\sl and} \address}

\def\abstract#1{\par\dimen@=\prevdepth \hrule height\z@ \prevdepth=\dimen@
   \vskip\frontpageskip\spacecheck\sectionminspace
   \centerline{\fourteencp ABSTRACT}\vskip\headskip
   {\noindent #1}}

\def\email#1{\fnote{\tentt e-mail: #1\hfill}}

%
%

%

%
\def\QMW{\address{%
   Department of Physics, Queen Mary and Westfield College\break
   Mile End Road, London E1 4NS, UK}}
%

%
%
\newcount\refnumber \def\clearrefnumber{\refnumber=0}  \clearrefnumber
\newwrite\R@fs                              
\immediate\openout\R@fs=\jobname.refs 
\def\closerefs{\immediate\closeout\R@fs} 
\def\refsout{\closerefs\refs
\unlockat
\input\jobname.refs
\lockat
\endrefs}
\def\refitem#1{\item{{\bf #1}}}
\def\ifundefined#1{\expandafter\ifx\csname#1\endcsname\relax}
\def\[#1]{\ifundefined{#1R@FNO}%
\global\advance\refnumber by1%
\expandafter\xdef\csname#1R@FNO\endcsname{[\the\refnumber]}%
\immediate\write\R@fs{\noexpand\refitem{\csname#1R@FNO\endcsname}%
\noexpand\csname#1R@F\endcsname}\fi{\bf \csname#1R@FNO\endcsname}}
\def\refdef[#1]#2{\expandafter\gdef\csname#1R@F\endcsname{{#2}}}
%
%
\newcount\eqnumber \def\cleareqnumber{\eqnumber=0}
\newif\ifal@gn \al@gnfalse  
\def\veqnalign#1{\al@gntrue \vbox{\eqalignno{#1}} \al@gnfalse}
\def\eqnalign#1{\al@gntrue \eqalignno{#1} \al@gnfalse}
\def\(#1){\relax%
\ifundefined{#1@Q}
 \global\advance\eqnumber by1
 \ifs@cd
  \ifs@c
   \expandafter\xdef\csname#1@Q\endcsname{{%
\noexpand\rm(\the\secnumber .\the\eqnumber)}}
  \else
   \expandafter\xdef\csname#1@Q\endcsname{{%
\noexpand\rm(\char\the\appnumber .\the\eqnumber)}}
  \fi
 \else
  \expandafter\xdef\csname#1@Q\endcsname{{\noexpand\rm(\the\eqnumber)}}
 \fi
 \ifal@gn
    & \csname#1@Q\endcsname
 \else
    \eqno \csname#1@Q\endcsname
 \fi
\else%
\csname#1@Q\endcsname\fi\global\let\@Q=\relax}
%
%
\newif\ifm@thstyle \m@thstylefalse
\def\mathstyle{\m@thstyletrue}
\def\proclaim#1#2\par{\smallbreak\begingroup
\advance\baselineskip by -0.25\baselineskip%
\advance\belowdisplayskip by -0.35\belowdisplayskip%
\advance\abovedisplayskip by -0.35\abovedisplayskip%
    \noindent{\caps#1.\enspace}{#2}\par\endgroup%
\smallbreak}
\def\m@kem@th<#1>#2#3{%
\ifm@thstyle \global\advance\eqnumber by1
 \ifs@cd
  \ifs@c
   \expandafter\xdef\csname#1\endcsname{{%
\noexpand #2\ \the\secnumber .\the\eqnumber}}
  \else
   \expandafter\xdef\csname#1\endcsname{{%
\noexpand #2\ \char\the\appnumber .\the\eqnumber}}
  \fi
 \else
  \expandafter\xdef\csname#1\endcsname{{\noexpand #2\ \the\eqnumber}}
 \fi
 \proclaim{\csname#1\endcsname}{#3}
\else
 \proclaim{#2}{#3}
\fi}
\def\Thm<#1>#2{\m@kem@th<#1M@TH>{Theorem}{\sl#2}}
\def\Prop<#1>#2{\m@kem@th<#1M@TH>{Proposition}{\sl#2}}
\def\Def<#1>#2{\m@kem@th<#1M@TH>{Definition}{\rm#2}}
\def\Lem<#1>#2{\m@kem@th<#1M@TH>{Lemma}{\sl#2}}
\def\Cor<#1>#2{\m@kem@th<#1M@TH>{Corollary}{\sl#2}}
\def\Conj<#1>#2{\m@kem@th<#1M@TH>{Conjecture}{\sl#2}}
\def\Rmk<#1>#2{\m@kem@th<#1M@TH>{Remark}{\rm#2}}
\def\Exm<#1>#2{\m@kem@th<#1M@TH>{Example}{\rm#2}}
\def\Qry<#1>#2{\m@kem@th<#1M@TH>{Query}{\it#2}}
%
%

%
\def\<#1>{\csname#1M@TH\endcsname}
%
%
\def\ref#1{{\bf [#1]}}
\def\ie{{\it i.e.\/}}
\def\nl{\hfil\break}
%
%

\def\lapprox{\hbox{\lower3pt\hbox{$\buildrel<\over\sim$}}}
\def\gapprox{\hbox{\lower3pt\hbox{$\buildrel<\over\sim$}}}
\def\quotient#1#2{#1/\lower0pt\hbox{${#2}$}}
\def\fr#1/#2{\mathord{\hbox{${#1}\over{#2}$}}}
\ifamsfonts
 \mathchardef\empty="0\hexmsb3F 
 \mathchardef\lsemidir="2\hexmsb6E 
 \mathchardef\rsemidir="2\hexmsb6F 
\else
 \let\empty=\emptyset
 \def\lsemidir{\mathbin{\hbox{\hskip2pt\vrule height 5.7pt depth -.3pt
    width .25pt\hskip-2pt$\times$}}}
 \def\rsemidir{\mathbin{\hbox{$\times$\hskip-2pt\vrule height 5.7pt
    depth -.3pt width .25pt\hskip2pt}}}
\fi
%
\def\to{\rightarrow}
\def\lra{\leftrightarrow}
%

%
%
\def\reals{\mathord{\bb R}} 
%
%
\def\underrightarrow#1{\vtop{\ialign{##\crcr
      $\hfil\displaystyle{#1}\hfil$\crcr
      \noalign{\kern-\p@\nointerlineskip}
      \rightarrowfill\crcr}}} 
\def\underleftarrow#1{\vtop{\ialign{##\crcr
      $\hfil\displaystyle{#1}\hfil$\crcr
      \noalign{\kern-\p@\nointerlineskip}
      \leftarrowfill\crcr}}}  

\def\comm#1#2{\left[#1\, ,\,#2\right]}
%
%
%
\def\PRL#1#2#3{{\sl Phys. Rev. Lett.} {\bf#1} (#2) #3}
\def\NPB#1#2#3{{\sl Nucl. Phys.} {\bf B#1} (#2) #3}

\def\CMP#1#2#3{{\sl Comm. Math. Phys.} {\bf #1} (#2) #3}
\def\PRD#1#2#3{{\sl Phys. Rev.} {\bf D#1} (#2) #3}

\def\PLB#1#2#3{{\sl Phys. Lett.} {\bf #1B} (#2) #3}

\def\PTP#1#2#3{{\sl Prog. Theor. Phys.} {\bf #1} (#2) #3}

\def\PR#1#2#3{{\sl Phys. Reports} {\bf #1} (#2) #3}

\def\FAaIA#1#2#3{{\sl Functional Analysis and Its Application} {\bf #1} (#2)
#3}

\def\JPA#1#2#3{{\sl J. Physics} {\bf A#1} (#2) #3}

\def\MPLA#1#2#3{{\sl Mod. Phys. Lett.} {\bf A#1} (#2) #3}

\def\JETPL#1#2#3{{\sl  Sov. Phys. JETP Lett.} {\bf #1} (#2) #3}

\lockat

\def\W{{\ssf W}}
\def\d{\partial}
\def\L{{\ssf L}}
\def\KdV{{\ssf KdV}}
\def\SKdV{{\ssf SKdV}}
\def\SBKP{{\ssf SBKP}}
\def\SKP{{\ssf SKP}}
\def\dddot#1{\hbox{$\mathop{#1}\limits^{\ldots}$}}
\def\ddxperp{\ddot x_\perp}

\def\dqperp{\dot q_\perp}

%
%
\refdef[reviews]{
K. Schoutens, A. Sevrin and P. van Nieuwenhuizen, \NPB{349}{791}{1991},\nl
C.M. Hull, \CMP{156}{245}{1993},\nl
J. De Boer and J. Goeree, \NPB{401}{369}{1993}, ({\tt
hep-th/9206098})\nl
S. Govindarajan and T. Jayaraman, {\sl ``A proposal for the
geometry of $W_n$ gravity''}, ({\tt hep-th/9405146}).}
\refdef[Polyakov]{A.M. Polyakov, \NPB{268}{406}{1986}.}
\refdef[Ambjorn]{J. Ambj\o rn, B. Durhuus and T. Jonsson,
\JPA{21}{981}{1988}.}
\refdef[Espriu]{F. Alonso and D. Espriu, \NPB{283}{393}{1987}.}
\refdef[extgeo]{G. Sotkov and M. Stanishkov, \NPB{356}{1991}{439};
G. Sotkov, M. Stanishkov and C.J. Zhu, \NPB{356}{1991}{245}.\nl
J.L. Gervais and Y. Matsuo, \PLB{274}{1992}{309} ({\tt hep-th/9110028});
\CMP{152}{1993}{317} ({\tt hep-th/9201026}).\nl
J.M. Figueroa-O'Farrill, E. Ramos and S. Stanciu,
\PLB{297}{1992}{289},
({\tt hep-th/9209002}).\nl
J. Gomis, J. Herrero, K. Kamimura and J. Roca, \PTP{91}{1994}{413}.}
\refdef[partcurv]{R.D. Pisarski, \PRD{34}{1986}{670}.\nl
M.S. Plyushchay, \PLB{253}{1991}{50}.\nl
C. Batlle, J. Gomis, J.M. Pons and N. Rom\' an-Roy,
\JPA{21}{1988}{2693}.}
\refdef[Plyus]{M.S. Plyushchay, \MPLA{4}{1989}{837}.}
\refdef[Radul]{A.O. Radul, \JETPL{50}{1989}{371}; \FAaIA{25}{1991}{25}.}
\refdef[GraciaPons]{X. Gr\`acia and J.M. Pons, \JPA{25}{1992}{6357}.}
\refdef[Dickey]{L.A. Dickey, \CMP{87}{1982}{127}.}
\refdef[Zoller]{D. Zoller, \PRL{65}{1990}{2236}.}
\refdef[Beckers]{K. Becker and M. Becker, \MPLA{8}{1205}{1993},
({\tt hep-th/9301017}).}
\refdef[JoseSonia]{J.M. Figueroa-O'Farrill and S. Stanciu,
\MPLA{8}{2125}{1993},
({\tt hep-th/9303168}).}
\refdef[nequalone]{J.M. Figueroa-O'Farrill and E. Ramos,
\CMP{145}{43}{1992}.}
\refdef[EduSonia]{E. Ramos and S. Stanciu, {\sl `` On the
supersymmetric BKP-hierarchy''},\nl
({\tt hep-th/9402056}).}
\refdef[Penrose]{R. Penrose and M. MacCallum, \PR{C6}{247}{1972}.}
\refdef[Zuminoetal]{L. Brink, S. Deser, B. Zumino, P. DiVecchia and
P.S. Howe, \PLB{64}{435}{1976}.}
\refdef[Friedan]{D. Friedan, E. Martinec and S. Shenker,
\NPB{271}{93}{1986}.}

%
\overfullrule=0pt
\def\pubblock{ \line{\hfil\rm QMW--PH--94-23}
               \line{\hfil\tt hep-th/9408019}
               \line{\hfil\rm July 1994}}
\titlepage
\title{$\W$-symmetry and the rigid particle}
\author{Eduardo Ramos\email{E.Ramos@qmw.ac.uk}}
\andauthor{Jaume Roca\email{J.Roca@qmw.ac.uk}}
\QMW
\abstract{We prove that $\W_3$ is the gauge symmetry of the
scale-invariant rigid particle, whose action is given by the
integrated extrinsic curvature of its world line.
This is achieved by showing that its equations
of motion can be written in terms of the Boussinesq operator.
The $\W_3$ generators $T$ and $W$ then appear
respectively as functions
of the induced world line metric and the extrinsic curvature.
We also show how the equations of motion for the standard
relativistic particle arise from those of the rigid particle
whenever it is consistent to impose the ``zero-curvature gauge'',
and how to rewrite them in terms of the $\KdV$ operator.
The relation between particle models and integrable systems is
further pursued in the case of the spinning particle, whose equations of
motion are closely related to the $\SKdV$ operator.
We also partially extend our analysis in the supersymmetric
domain to the scale invariant rigid particle by explicitly constructing
a supercovariant version of its action.}
\endtitlepage
\section{Introduction}

Since their appearance in the seminal work of A.B. Zamolodchikov,
$\W$-algebras have played a central role in conformal field
theory and string theory. Nevertheless, many of their properties
are not yet fully understood mainly due to their nonlinear character.
In particular, a complete understanding of their underlying geometry
is still lacking, though many partial results are already avalaible
\[extgeo]\[reviews].
It is clear that simple mechanical systems displaying $\W$ symmetry
could be an invaluable tool in this difficult task. On the one hand, they
could provide us with some geometrical and/or physical interpretations
for $\W$-transformations ($\W$-morphisms), while on the other hand
they could give us some hints as to which structures are those
associated with $\W$-gravity -- the paradigmatic example being
provided by
the standard relativistic particle and diffeomorphism invariance
($\W_2$).

The connection between $\W$-morphisms and the extrinsic geometry of
curves and surfaces is well known by now \[extgeo].
Therefore, it seems natural to look
for a $\W$-particle candidate among the geometrical actions depending
on the extrinsic curvature. The canonical analysis of those models
has been carried out by several authors \[partcurv]. In particular
M.S.~Plyushchay studied in \[Plyus] the action given by
$$S=\alpha \int \sqrt{|\kappa^2|} \; ds, \(action)$$
where $\alpha$ is a dimensionless coupling constant\fnote{In our convention
the coordinates are dimensionless.} and the extrinsic curvature
$\kappa$ is given by
$$\kappa^2 = g_{\mu\nu}{d^2 x^{\mu}\over {d s^2}}
{d^2 x^{\nu}\over {d s^2}},\(curvature)$$
where $g_{\mu\nu}$ stands for a minkowskian, euclidean, or any
$x$-independent metric.
He showed that this dynamical system
possesses two gauge invariances, and that one of them is
the expected invariance under diffeomorphisms. The main
purpose of this paper is to show that these gauge invariances
are nothing but $\W_3$.

Before entering into more technical matters, we would like to point out
that the scale invariant rigid particle plays an important role
in such diverse areas as polymer physics, random surfaces
and string theory. In particular, since Polyakov pointed out in
\[Polyakov] the
relevance of scale invariant extrinsic curvature terms in the string
action for a possible stringy description of QCD, the rigid particle
has been used as a simpler laboratory where some of these ideas could
be tested \[Ambjorn]\[Espriu]. In view of this, we believe that the
connection of this system with $\W$-symmetry is interesting
from both the physical and mathematical point of view, and that
an understanding of $\W$-geometry can find direct application in
models of physical relevance.

The plan of the paper is as follows. In section 2 we review some
well-known results about the scale invariant rigid particle that
will be useful for our purposes.

In section 3 we present the main results of the paper. It is shown
how the equation of motion associated with the
scale invariant rigid particle can be recast in the Boussinesq form, and
how $\W_3$-invariance follows from this. We also show how the
equation of motion of the standard relativistic particle is obtained
in the ``zero curvature'' gauge, and we comment on the appearence of
the $\KdV$ operator in this case.

In section 4 we extend the connection between particle models
and integrable systems to the supersymmetric domain. The reason to do
so, besides that of completeness, is because there is no
unique supersymmetrization of the Lax operators associated with
generalized $\KdV$-hierarchies
\[Beckers]\[JoseSonia], and therefore we expect that
these models can shed some light on the physical
significance of these different supersymmetrizations.
We then show that the equation of motion of
the standard spinning particle
can be recast in terms of the first superLax operator of the
$N=1$ series of \[nequalone]. We also construct the supersymmetric
version of the scale invariant rigid particle. Unfortunately, the
expression of the action is quite involved, even when written in
terms of superfields, and we have been unable to
explicitly display the relationship between its equations of motion
and any superintegrable model (although we have no doubt of its
existence).

We finish with some comments about possible generalizations of this
formalism to $\W_n$ with $n\geq 4$, and we speculate about the
possible relevance of our results for the construction of a fully covariant
$\W_3$-gravity theory.
\vskip 0.5truecm

\section{Hamiltonian analysis of the rigid particle}

We now briefly review some standard results concerning the gauge
structure of the action \(action) which will be useful in the following.

In an arbitrary parametrization $x^\mu(t)$ the lagrangian
is given by
$$
L=\alpha\sqrt{\left|{\ddxperp^2}\over{\dot x^2}\right|},\(lagran)
$$
where $\dot x^\mu=dx^\mu/dt$ and $\ddxperp^\mu=\ddot x^\mu
-\dot x^\mu(\ddot x\dot x)/\dot x^2$.

For the study of the system in the hamiltonian framework
it is convenient to introduce the velocity $q^\mu=\dot x^\mu$
as an auxiliary variable in order to avoid the presence of
second-order derivative terms in the lagrangian. The
lagrangian in these new variables reads
$$L =\alpha\sqrt{\left|{\dqperp^2}\over{q^2}\right|} + \gamma
(q - \dot x).\(constlagran)$$

We introduce the canonical momenta $(P_\mu,p_\mu,\pi_\mu)$
associated with the coordinates $(x^\mu,q^\mu,\gamma^\mu)$.
Their Poisson bracket algebra is given by
$$
\{{\cal Q}^\mu,{\cal P}_\nu\}=\delta^\mu_{\;\nu}.\(poissonbra)
$$
The definition of the momenta implies a number of primary hamiltonian
contraints.
Being the lagrangian linear in $\dot x$ we obtain the trivial
second-class constraints $\gamma^\mu=-P^\mu$ and $\pi^\mu=0$,
from which we can eliminate the canonical pair $(\gamma,\pi)$
by means of the Dirac bracket.
The primary first-class contraints are
$$
\phi_1=pq\approx0,\quad\quad\phi_2={{1}\over{2}}\left(|p^2|-{{\alpha^2}\over
{|q^2|}}\right)\approx0.
\(primfcconst)
$$
Stabilization of these constraints through the standard Dirac
procedure leads to the secondary
$$
\phi_3=Pq\approx0,\quad\quad\phi_4=Pp\approx0,
\(secfcconst)
$$
and tertiary
$$
\phi_5=P^2\approx0,
\(tertfcconst)
$$
first-class hamiltonian constraints, which form a closed
Poisson bracket algebra.

In the constrained submanifold, time-evolution is generated
by the hamiltonian
$$
H=\phi_3+v_1\phi_1+v_2\phi_2.
\(hamiltonian)
$$
The functions $v_1$ and $v_2$ have a definite expression in terms
of lagrangian quantities,
$$
v_1={{q\dot q}\over{q^2}},\quad\quad
v_2={{1}\over{\alpha}}\sqrt{|q^2\dqperp^2|},
\(vi)
$$
but they should be regarded as arbitrary
functions of time in the canonical formalism. Time evolution
becomes unambiguous once we assign to them definite values, which is
nothing but a choice of gauge.
The hamiltonian stabilization procedure guarantees the consistency
of the expressions chosen for $v_1$ and $v_2$ with their lagrangian
expression \(vi).

The presence of two arbitrary
functions reveals the existence of a second gauge symmetry in addition
to the familiar reparametrization invariance.
We shall show in what follows that their symmetry algebra is
precisely $\W_3$.
\vskip 0.5truecm

\section{ Equations of motion, Boussinesq operator and
$\W_3$ symmetry}

The hamiltonian equations of motion can be suggestively written as
$$\eqalign{
\dot x^\mu&=q^{\mu},
\cr
\left(\matrix{
\dot P^\mu\cr
\dot p^\mu\cr
\dot q^\mu\cr}\right) & =
\left(\matrix{0&0&0\cr
-1&-v_1&-{\alpha^2 v_2\over {q^4}}\cr
0&v_2&v_1\cr}\right)
\left(\matrix{P^\mu\cr p^\mu\cr q^\mu\cr}\right).
}\(eqmotion)$$
Notice that, neglecting the first equation,
which simply states the definition of $q^{\mu}$, the equations of
motion can be casted in the ``Drinfeld-Sokolov'' form
associated with a particular $SL(3)$ connection. We
will come back to this point later on.

It will be convenient in what follows to introduce a different and
more geometrical parametrization for $v_1$ and $v_2$
Let us define $e$ and $\lambda$ as
$$
e^2=
|q^2|,\quad\quad\quad\lambda^2=|\kappa^2|.\(repara)
$$

Notice that $e^2$ is nothing
but the modulus of the induced metric, and $\lambda$
is basically the extrinsic curvature.
In terms of $e$ and $\lambda$
the gauge degrees of freedom are given by $v_1 = \dot e/e$ and
$v_2 =\lambda e^3/\alpha$.

{}From \(eqmotion) we can obtain a single equation for the velocity
vector $q^\mu$,
$$
\dddot q^\mu+u_1\ddot q^\mu+u_2\dot q^\mu+u_3q^\mu=0,
$$
with
$$\eqalign{
u_1&=-2{{d\ln(e^3 \lambda )}\over{dt}},
\cr
u_2&={e^4\lambda^2}+15{{\dot e^2}\over{e^2}}+7{{\dot e\dot
\lambda}\over{e\lambda}}+2{{\dot\lambda^2}\over{\lambda^2}}
-4{{\ddot e}\over{e}}-{{\ddot\lambda}\over{\lambda}},
\cr
u_3&=e^4\dot\lambda\lambda + e^3\dot e\lambda^2
-15{{\dot e^3}\over{e^3}}-7{{\dot e^2\dot\lambda}\over{e^2\lambda }}
-2{{\dot e\dot\lambda^2}\over{e\lambda^2}}
+10{{\dot e\ddot e}\over{e^2}}+2{{\ddot e\dot\lambda}\over{e\lambda}}
+{{\dot e\ddot\lambda}\over{e\lambda}}-{{\dddot e}\over{e}}.\cr
}\(choris)$$
Notice that $u_1$ is a total derivative. This
allows us to remove the term with the second derivative by a simple
local rescaling of $q^\mu$. Indeed, if we define
$$
y^\mu={1\over e^2\lambda^{2\over 3}}q^\mu,\(redefinition)
$$
the equation for the new velocity vector $y^\mu$ can be written
in terms of the Boussinesq Lax operator
$$
\dddot y^\mu+T\dot y^\mu+(W+{{\dot T}\over2})y^\mu=0,\(Boussequation)
$$
with $T$ and $W$ given by
$$
\eqalign{
T=&\; {e^2\lambda^2}-{1\over3}{\dot e^2\over e^2}-
{\dot e\dot\lambda\over e\lambda}-{4\over3}{\dot\lambda^2\over
\lambda^2}+ 2{\ddot e\over e}+{\ddot\lambda\over\lambda},
\cr
W=&\; e \dot e \lambda^2 + 3 {\dot e^3\over e^3} +{5\over 3}
e^2 \lambda\dot\lambda + {\dot e^2\dot\lambda\over e^2\lambda} +
{4\over 3}{\dot e\dot\lambda^2\over e \lambda^2} +{56\over 27}
{\dot\lambda^3\over\lambda^3}\cr
&-4{\dot e \ddot e\over e^2} -{2\over 3}{\ddot e\dot\lambda\over e
\lambda}-{\dot e\ddot\lambda\over e\lambda} -{8\over 3}
{\dot\lambda\ddot\lambda\over \lambda^2} + {\dddot e\over e}+
{2\over 3}{\dddot\lambda\over\lambda}.
}$$

The algebra of symmetries of equations of the type $\L \Psi =0$ with
$\L$ a differential operator of the form $\partial ^n + ...$ has been
studied by Radul in \[Radul]. From his general construction it is easily
deduced that for the particular case of $\L = \d^3 + T\d + W + \dot
T/2$, which is our case of interest, the symmetry algebra is $\W_3$.
Rather than reproducing here the general arguments leading to this
result, we will work out the case at hand explicitly.

The most general (local) variation of $\Psi$ preserving the
structure of $\L$ , up to
equations of motion, is given by
$$\delta_{\epsilon ,\rho}\Psi =
\rho \ddot\Psi + (\epsilon -{1\over 2}\dot\rho )\dot\Psi
-(\dot\epsilon +{1\over 6}\ddot\rho + {2\over 3} \rho T)\Psi,
\(varpsi)$$
where the parametrization has been chosen so that $\epsilon$ and
$\rho$ denote the parameters associated with diffeomorphisms and
pure $\W_3$-morphisms respectively. The corresponding transformations
for $T$ and $W$ which leave the equations of motion invariant are
given by
$$\eqalign{
\delta^{(T)}_{\epsilon}T = & \; 2 \dddot\epsilon + 2 \dot\epsilon T
+\epsilon\dot T,\cr
\delta^{(T)}_{\epsilon}W = & \; 3\dot\epsilon W +\epsilon\dot W,\cr
\delta^{(W)}_{\rho} T = &\; 3 \dot\rho  W + 2\rho \dot W,\cr
\delta^{(W)}_{\rho} W = & \; -{1\over 6} \rho^{({\rm v})}
- {5\over 6} \dddot\rho
T -{5\over 4}\ddot\rho\dot T - \dot\rho ({3\over 4}\ddot T
+{2\over 3}T^2) -\rho ({1\over 6} \dddot T +{2\over 3} T\dot T).\cr
}\(wmorphisms)$$

It is now a long but straightforward exercise to check that these
transformations obey the algebra of $\W_3$ transformations, {\ie}
$$\eqalign{
\comm{\delta^{(T)}_{\epsilon_1}}{\delta^{(T)}_{\epsilon_2}} = & \;
\delta^{(T)}_\epsilon,\quad{\rm where}\quad
\epsilon=\epsilon_1\dot\epsilon_2 -\dot\epsilon_1\epsilon_2,
\cr
\comm{\delta^{(T)}_{\epsilon}}{\delta^{(W)}_{\rho}} = & \;
\delta^{(W)}_{\tilde\rho},\quad{\rm where}\quad
\tilde\rho=\epsilon\dot\rho - 2\dot\epsilon\rho,
\cr
\comm{\delta^{(W)}_{\rho_1}}{\delta^{(W)}_{\rho_2}} = & \;
\delta^{(T)}_\epsilon,\quad{\rm where}\quad
\epsilon={{2\over 3}\rho_1\dot\rho_2 T -{1\over 4}\dot\rho_1
\ddot\rho_2 +{1\over 6}\rho_1\dddot\rho_2 - (\rho_1\lra\rho_2)},\cr
}\(commutators)$$
where we have assumed for simplicity that the parameters are field
independent. The most general case can be equally worked out from
the results of \[Radul].

Notice that the consistency of the procedure in our case
is based on the fact that
$T$ and $W$, through their dependence in $\lambda$ and $e$, are gauge
degrees of freedom themselves, therefore before any solution to the
equations is found they should be given definite values. In this
language the $\W$-morphisms given by \(wmorphisms) are to be
understood as gauge transformations.

\subsection{$\W$-symmetry as an invariance of the action}

Being that the relation between $y^\mu$ and $q^{\mu}$  is not
easily invertible, the previous analysis does not yield
the transformation rules for $q^\mu$ directly. These transformations
are interesting by themselves because the action is naturally
written in terms of $q^\mu$. Here is where standard techniques
in constrained dynamical systems come to our rescue.
The gauge transformations for $x^\mu$, $q^\mu$, and $\gamma^\mu$
can be computed
using the general method of \[GraciaPons].
We shall give here a brief outline of the procedure.

Let us introduce an operator $K$ which acts on arbitrary functions
$g({\cal Q},{\cal P},t)$ in phase space, giving (on-shell) their
time-derivative in velocity space:
$$
K\cdot g=\dot{\cal Q}\widehat{\left({{\partial g}\over{\partial{\cal Q}}}
\right)}+{{\partial L}\over{\partial{\cal Q}}}\widehat{\left({{\partial g}
\over{\partial{\cal P}}}\right)}+\widehat{\left({{\partial g}\over{\partial
t}}\right)}.
$$
where $\widehat h({\cal Q},\dot{\cal Q})$ stands for the function in velocity
space
obtained from $h({\cal Q},{\cal P})$ after substituting the momenta by
their lagrangian expression: ${\cal P}=\partial L/\partial\dot{\cal Q}$.

The operator $K$ plays a relevant role in the construction of gauge
transformations. In particular, it can be shown that all lagrangian
constraints are obtained by applying $K$ to the hamiltonian constraints.

Let us consider the submanifold $V_1$ defined by the primary lagrangian
constraints $\chi^1_a=K\cdot\phi^0_a$, where $\phi^0_a$ are the primary
hamiltonian constraints.
It can be shown \[GraciaPons] that a hamiltonian function $G$ satisfying
$K\cdot G\approx0,$ on $V_1$, \ie
$$
K\cdot G+\sum_ar^a\chi^1_a=0,
$$
can be used to construct Noether gauge transformations
in the following way:
$$
\delta {\cal Q}=\widehat{\left({{\partial G}\over{\partial{\cal P}}}\right)}
+\sum_ar^a\gamma_a,
\(Noethertransf)
$$
where $\gamma_a$ form a basis for the null vectors of the hessian
matrix $\partial^2 L/\partial{\cal Q}^2$ and have the
expression
$$
\gamma_a=\widehat{\left({{\partial\phi^0_a}\over{\partial{\cal P}}}\right)}.
$$

In practice, it is enough to consider $G$ as a linear combination of
hamiltonian constraints with arbitrary infinitesimal functions as their
coefficients.
Imposing $K\cdot G\approx0$ on $V_1$ determines some of these functions in
terms of the rest. Those that remain arbitrary are to be considered as the
infinitesimal parameters of the gauge transformations.
Primary hamiltonian costraints can, in fact, be excluded from the previous
combination as they have no contribution in the gauge transformations.

In the case at hand the `generator' $G$ and the functions $r^a$ are given by
$$\eqalign{
G &=\beta(t)\phi_3+\dot\eta(t)\phi_4+\eta(t)\phi_5,
\cr
r^1 &=\dot\beta+{\dot e\over e}\beta-\alpha{\lambda\over e}\dot\eta,
\cr
r^2 &={1\over\alpha}e^3\lambda\beta+\ddot\eta-{\dot e\over e}\dot\eta,}
$$
and from \(Noethertransf) we obtain the gauge transformations:
$$\eqalign{
\delta_{\beta,\eta}x^\mu &=\beta q^\mu +
\dot\eta {\alpha\over e^3 \lambda}\dot q^\mu_{\perp}- 2\eta\gamma^\mu,\cr
\delta_{\beta,\eta}q^\mu &= \dot\beta q^\mu + \beta\dot q^\mu
-\dot\eta\left( \gamma^\mu +\alpha {\lambda\over e} q^\mu +
\alpha {\dot e\over e^4\lambda}\dot q^\mu_{\perp}\right) +
\ddot\eta {\alpha \over e^3\lambda}\dot q^\mu_{\perp},\cr
\delta_{\beta,\eta}\gamma^\mu &=0.\cr
}\(varq)$$
A direct computation now shows
that the tranformations given by \(varq) are not
only a symmetry of the action, but also that, as a soft algebra,
reproduce the algebra of $\W_3$ transformations if we make the following
identifications:
$$\eqalign{
\epsilon &=\beta + {\alpha\over 2}{\ddot\eta\over e^3 \lambda}
+\dot\eta {\alpha\over e^3\lambda}\left( {1\over 6}{\dot\lambda
\over \lambda} - {1\over 2}{\dot e\over e}\right)\cr
\rho &= - {\alpha\over e^3\lambda}\dot\eta.\cr}\(changepara)$$
Notice that in this parametrization the variation of $x^\mu$ has a
nonlocal expression. This is not entirely unexpected due to the
fact that the natural variable for the system seem to be supplied by
$q^\mu$ rather than the coordinates themselves\fnote{This is
somehow reminiscent of what happens for the case of the two
dimensional massless boson, where the natural variable is supplied
by its associated $U(1)$ current.}.
Moreover, as expected,
the variations induced on $y^\mu$ through \(varq)
coincide with the ones obtained via \(varpsi), which tightens up the
formalism, and shows the nice interplay between the hamiltonian
and the $\W$-algebraic methods of tackling the problem.

There is also a more indirect, but also interesting, way to
show the invariance of the action \(action) under $\W_3$-morphisms
with the help of the Miura transformation. In order to do so we
should return to the expression of the hamiltonian equations of
motion in terms of the $SL(3)$ connection. If we write \(eqmotion) as
$$\left( {{d\ }\over {dt}} - \Lambda \right)\left(\matrix{P^\mu\cr
p^\mu\cr q^\mu\cr}\right)=0,\(DrinSok)$$
it is clear that these equations have local gauge invariance
under
$$\left(\matrix{P^\mu\cr p^\mu\cr q^\mu\cr}\right)\to
M \left(\matrix{P^\mu\cr p^\mu\cr q^\mu\cr}\right)\quad
{\rm and}\quad \Lambda\to M\Lambda M^{-1} - M { dM^{-1}\over dt },$$
with $M\in SL(3)$.
It will be convenient for our purposes to work in the Miura gauge,
{\ie}
$$\Lambda = \left(\matrix{-\varphi_1 -\varphi_2&0&0\cr
1&\varphi_2&0\cr 0&1&\varphi_1}\right).\(miuragauge)$$
The matrix $M\in SL(3)$ which brings $\Lambda$ to this form is
$$M ={\lambda^{1\over 3}\over\alpha^{1\over 3}}
\left(\matrix{e&0&0\cr
0&-e&-i {\alpha}/e\cr
0&0& -\alpha/e^2\lambda}\right).$$
Some straightforward algebra now yields
$$\eqalign{\varphi_1 &= -{\dot e\over e} -{2\over 3}{\dot\lambda
\over\lambda} - i {\lambda e}\cr
\varphi_2 &= {1\over 3}{\dot\lambda \over\lambda}
+ i {\lambda e}.\cr}\(miurafields)$$
The key point, for our present interest, is given by the fact that
the lagrangian \(lagran) can be written as
$$L \sim {\rm Im}(\varphi_1 ) \sim {\rm Im}(\varphi_2 ).\(lagmiura)$$
The transformation under $\W_3$-morphisms of the Miura fields can
be computed using the Kupershmidt-Wilson
theorem \[Dickey]. If we denote by $\epsilon$
the parameter associated with diffeomorphisms, and by $\rho$ the
one associated with pure $\W_3$-morphisms, these variations
are given by
$$\eqalign{\delta_{\epsilon ,\rho }\varphi_1 =
{{d\ }\over dt} & \left( \dot \epsilon + \epsilon\varphi_1 +
{1\over 6} \ddot\rho +{1\over 2}(\dot\varphi_1\rho
+\varphi_1\dot\rho)\right.
\cr
& \left.-{1\over 3}({1\over 2}\dot\varphi_1 -2\dot\varphi_2 -\varphi_1^2 + 2
\varphi_2^2 + 2\varphi_1\varphi_2 )\rho\right),\cr
\delta_{\epsilon ,\rho}\varphi_2 = {{d\ }\over dt} &
\left( \epsilon\varphi_2 -
{1\over 3} \ddot\rho -{1\over 2}(\dot\varphi_2\rho +\varphi_2\dot\rho
) -\dot\varphi_1\rho +\varphi_1\dot\rho\right.
\cr & \left.
+{1\over 3}(\dot\varphi_1 +{1\over 2}\dot\varphi_2 - 2\varphi_1^2 +
\varphi_2^2 - 2\varphi_1\varphi_2 )\rho\right),\cr
}\(transfos)$$
which is a total derivative! And from this follows directly the
invariance of the action.

\subsection{The relativistic particle}

It is easy to recover the equations of motion for the standard
relativistic particle from \(eqmotion) whenever it is consistent to
impose the gauge condition $v_2 =0$. Notice that since $v_2$
is proportional to the extrinsic curvature, this
gauge can only be consistently imposed when it can be taken to be zero
(initial conditions may be incompatible with this gauge choice).
In that case the
hamiltonian equations of motion collapse to
$$ \ddot x^\mu - {\dot e\over e}\dot x^\mu =0,\(relpartequ)$$
which is the equation of motion for the relativistic particle.
Notice that \(relpartequ) can be sugestively written
as $\ddot x^\mu_{\perp} = 0$.

It is natural to ask now, in the light of our previous discussion,
whether the $\KdV$ Lax operator appears in this
case. The answer to this question is easily obtained by realising
that via the redefinition
$$ x^\mu\to e^{{1\over 2}}x^\mu\,\(redefpart)$$
the equations of motion for the new variable can be written as
$$\ddot x^\mu + T x^\mu =0,\(kdvform)$$
with
$$T={1\over 2}{\ddot e\over e} -{3\over 4}{\dot e^2\over e^2},
\(Tpar)$$
and $e$ being a gauge degree of freedom for the relativistic
particle we can apply all the same arguments as above.
\vskip 0.5truecm

\section{The supersymmetric arena}

We shall now show how some of our previous results can be extended to the
supersymmetric domain.

\subsection{The spinning particle and \SKdV}

Since the spinning particle provides the natural world-line supersymmetrization
of the relativistic particle it is natural to address the question of which
supersymmetric Lax operator, if any, appears in this case.

The massive spinning particle is described by the action \[Zuminoetal]
$$
S={1\over 2}\int dt\left({{\dot x^2}\over{e}}+m^2e-\psi\dot\psi+\beta\dot\beta
-{\chi\over e}\psi\dot x+m\chi\beta\right),
\(spinaction)
$$
where $\psi^\mu$ is the supersymmetric partner of the particle
position $x^\mu$,
$\chi$ is the world line gravitino and $\beta$ is an auxiliary
fermionic variable which is necessary to introduce in order
to deal with the massive case.

This model is invariant under $N=1$ superdiffeomorphisms \[Friedan]. This can
be made explicit by writing its action in terms of superfields. Let us
first introduce the
superfields $X^\mu$ and $E$, of weights 0 and 1, associated with
the coordinate $x^\mu$ and the einbein $e$ respectively:
$$
X^\mu=x^\mu+\theta e^{1/2}\psi^\mu,\quad\quad
E=e+\theta e^{1/2}\chi.
\(superfielduno)
$$
In contrast, we do not have among the initial fields a natural
superpartner of $\beta$.
In fact, one can show that supersymmetry transformations close on
$\beta$ only on-shell. This problem can be circumvented by introducing
a new auxiliary
variable $b$ and its associated zero-weight superfield,
$$
B=\beta +\theta b
\(superfielddos)
$$

In terms of these three superfields the action can be written in the form
$$
S={1\over2}\int dt\;d\theta\left(E^{-1}\;DX D^2X -BDB+2mBE^{1/2}\right).
\(superactionsp)
$$
Notice that even though $D^2X^\mu$ is not a tensor superfield, the
nontensorial term in its transformation under superdiffeomorphisms
is proportional to $DX^\mu$, thus rendering the action invariant.

The matter equations of motion can be easily computed to be
$$
D^3X^\mu-{1\over2}E^{-1}DED^2X^\mu-{1\over2}E^{-1}D^2EDX^\mu=0.
\(sp eq mo)
$$
If in analogy with the bosonic case we now perform the change of variables
$$
X^\mu\rightarrow E^{1/2}X^\mu,
\(superredef)
$$
we can eliminate the second order term in \(sp eq mo) and the equations of
motion read
$$
D^3X^\mu+UX^\mu=0,
\(skdv eq)
$$
with
$$
U={1\over2}\left({{D^3E}\over E}-{3\over2}{{DED^2E}\over{E^2}}\right).
\(supert)
$$
It is remarkable that the first order term also disappears so that
equation \(skdv eq)  is precisely the first superLax operator associated
with the $N=1$ series of \[nequalone].
It was shown in \[EduSonia] how this $N=1$ series
provide us with hamiltonian structures for the $\SBKP$ reduction of the
$\SKP_2$ hierarchy. In particular, the operator $D^3 + U$ provides us
with a hamiltonian structure for the $\SKdV$ hierarchy, thus making
explicit the connection between the spinning particle and
superintegrability.

\subsection{The spinning scale-invariant rigid particle}

In order to perform the world line supersymmetrization of the
scale invariant rigid particle it will prove convenient to rewrite
its action \(action) with the help of two auxiliary fields $e$ and
$\lambda$ as follows
$$S=\alpha\int dt \left( e\lambda + {1\over 2}{\ddot x^2_{\perp}\over
e^3\lambda} - {1\over 2}{\dot x^2\over e}\lambda \right).\(newaction)$$

The equations of motion for $e$ and $\lambda$ imply that
their explicit expressions coincide with the ones given in \(repara),
thus justifying their
name. Moreover, since their equation of motion are algebraic we can
substitute them back in \(newaction) and recover
the usual form of the action \(action).

It is now simple with the experience of the standard relativistic
particle to attempt the supersymetrization of \(newaction). First
of all,
we should introduce an even superfield $\Lambda$ associated with $\lambda$,
{\ie} $\Lambda = \lambda +\theta \eta$. But this is not
enough and we should find an explicit expression
of the perpendicular projector preserving
supercovariance. This can be readily done by observing that
$$\Gamma = {1\over 2} {D\Omega\over \Omega},\(superconnection)$$
with
$$\Omega = (D^2 X)^2 + 2 D X D^3 X,\(omega)$$
transforms under superdiffeomorphisms as a superconnection.
Therefore we can define a supercovariant derivative,
$\nabla_{\Gamma}$, acting on
superfields $\Phi_h$, of weight $h$, with the help of $\Gamma$ by
$$\nabla_{\Gamma}\Phi_h = D\Phi_h - h\Gamma\Phi_h.\(nabla)$$

With all of this in mind we can directly  check
that the action given by
$$\alpha\int dt d\theta \left( {1\over 2}
{\nabla_{\Gamma}^4 X\nabla_{\Gamma}^3 X\over
E^3 \Lambda} - {1\over 2}{D X D^2 X\over E}\Lambda
-B DB +2 B( E K )^{1\over 2}\right),\(superaction)$$
is explicitly supercovariant, while reproducing \(newaction) when we
set the superpartners to zero. Unfortunately, when
the supercovariant derivatives
are expanded in terms of the superderivatives of $X^{\mu}$
the expression obtained is so
involved that has prevented us from making explict the connection with
superintegrability.
\vskip 0.5truecm

\section{Final comments and some conjectures}

We hope to have convinced the reader that $\W_3$-symmetry does play
an important role in the symmetry structure of the
rigid particle, and that standard techniques in constrained hamiltonian
systems and $\W$-algebras can be intertwined as powerful tools for
the better understanding of both disciplines.
It is also quite clear from our results that the connection between
integrability and particle models does not restrict to the rigid
particle case and seems to be present in other physically relevant
particle models.
But, of course, much is still
to be done. For example, we do not yet completely understand the
appearance of Lie algebraic structures in the problem, which should be
the key to generalize these procedures to other $\W$-algebras.
Although it seems plausible that a particle model with action
depending on the $(n-1)$-order extrinsic curvature will enjoy $\W_n$
symmetry, explicit computations become untractable and a more powerful
machinery should be developed.

Under quantization \[Plyus] the rigid particle is associated
with massless representations of the Poincare group
with integer helicity, therefore becoming a potential candidate for
a particle description of photons, gravitons and higher spin fields.
It is a tantalizing possibility that $\W_3$ symmetry can play a role
in such physical systems. In fact there are some indications that
it is indeed the case. It is clear from the transformation rules under
pure $\W_3$-morphisms that the world line element of the particle has
no gauge invariant meaning\fnote{This is of course a well-known fact,
although to the best of our knowledge, the gauge transformations which
render the position vector a gauge variant object were not displayed in
the literature.}. This nicely connects with some general arguments
\[Penrose] implying that it is impossible to assign, even classicaly,
a position to a single photon (or similarly, any massless
higher spin fields).

We would like to finish with a few sentences on the possible
relevance of our results to $\W_3$-gravity. It was shown in \[Zoller]
that a naive coupling to gravity of the action \(action), {\ie}
considering arbitrary $x$-dependent metrics, changes the gauge structure of
the system, and only reparametrization invariance survives.
This is due to the fact that the constraint algebra
no longer closes in curved space-time. We believe this to be a signal
indicating that the rigid particle is only consistently coupled to
$\W_3$-gravity.
Let us elaborate on this.
In the standard particle case, the relevant bundle is the tangent
bundle of the manifold, $TM$. In this case, it is well known how to equip
the bundle with a metric structure and all the powerful and
well-understood machinery of Riemmanian geometry is at our disposal.
In the language of jet bundles $TM$ is nothing but $J^1(\reals , M)$, and
it is our understanding that the relevant bundle for the $\W_3$-particle
is provided by $J^2(\reals ,M)$, which is not itself a vector bundle.
It is well known that $J^2(\reals , M)$ is an associated bundle to
$F^2M $, the frame bundle of second order, and it is our belief that
the required structure should be a ``natural'' structure in $F^2 M$,
as it is the metric in $FM$. It is our hope that the action of the
rigid particle will provide us with a ``$\W_3$-line element'', thus
offering some valuable insight about which generalized
structures in $F^2M$ should be considered.
Work on this is in progress.
\vskip 0.5truecm
\ack
We would like to thank J.M. Figueroa-O'Farrill and C.M. Hull for many useful
discussions on the subject.  J.R. is also grateful to the Spanish ministry of
education and the British council for financial support.
\vskip 0.5truecm

\refsout
\bye